\documentclass[10pt]{article}

\usepackage[T1]{fontenc}
\usepackage[utf8]{inputenc}
\usepackage{lmodern}
\usepackage{microtype}
\usepackage[margin=0.85in]{geometry}
\usepackage{amsmath,amssymb}
\usepackage{xcolor}
\usepackage{listings}   
\usepackage{booktabs,array,longtable,calc}
\usepackage{caption}
\usepackage{fvextra}
\usepackage{authblk}
\usepackage{titlesec}
\usepackage[numbers,sort&compress]{natbib}
\usepackage[hidelinks]{hyperref}
\usepackage{xurl}
\usepackage{booktabs}
\usepackage{multirow}
\usepackage{parskip}
\usepackage{listings}
\usepackage{xcolor}

\definecolor{codebg}{RGB}{248,248,248}
\definecolor{codecomment}{RGB}{95,135,95}
\definecolor{codekeyword}{RGB}{0,92,170}
\definecolor{codestring}{RGB}{170,70,40}
\definecolor{codenumber}{RGB}{120,120,120}

\lstdefinelanguage{Julia}{
  morekeywords={
    abstract,break,case,catch,const,continue,do,else,elseif,end,export,
    false,finally,for,function,global,if,import,in,let,local,macro,module,
    mutable,primitive,quote,return,struct,true,try,using,where,while
  },
  sensitive=true,
  morecomment=[l]\#,
  morestring=[b]",
  morestring=[b]'
}

\lstdefinestyle{julia}{
  language=Julia,
  backgroundcolor=\color{codebg},
  basicstyle=\ttfamily\small,
  keywordstyle=\color{codekeyword}\bfseries,
  commentstyle=\color{codecomment}\itshape,
  stringstyle=\color{codestring},
  numberstyle=\tiny\color{codenumber},
  numbers=none,
  frame=single,
  rulecolor=\color{black!15},
  framesep=5pt,
  xleftmargin=4pt,
  xrightmargin=4pt,
  aboveskip=6pt,
  belowskip=6pt,
  showstringspaces=false,
  keepspaces=true,
  columns=fullflexible,
  breaklines=true,
  breakatwhitespace=false,
  tabsize=4
}

\title{\texttt{ESRIcascade.jl}: A Julia package for fast computation of the Economic Systemic Risk Index}
\author[1,2]{Mitja Devetak}
\author[1,2,3,4]{Peter Klimek}
\affil[1]{Complexity Science Hub Vienna, Vienna, Austria}
\affil[2]{Supply Chain Intelligence Institute Austria, Vienna, Austria}
\affil[3]{Section for the Science of Complex Systems, Center for Medical Data Science, Medical University of Vienna, Vienna, Austria}
\affil[4]{Division of Insurance Medicine, Karolinska Institutet, Department of Clinical Neuroscience, Karolinska, Sweden}

\begin{document}
\maketitle

\begin{abstract}
This work presents \texttt{ESRIcascade.jl}, a Julia \cite{bezanson2017julia} package for computing the Economic Systemic Risk Index (ESRI) \citep{diem2022esri}. The package combines computational efficiency with a documented interface designed for research workflows. Our benchmarks show that \texttt{ESRIcascade.jl} is from 140 to 300 times faster than existing implementations. For a realistically sized synthetic economy, this reduces computational time from 16 hours to 7 minutes. By reducing computation times, \texttt{ESRIcascade.jl} makes repeated analyses of larger production networks practical and feasible and may enable broader use of ESRI in empirical and methodological research. The package and its documentation are freely available at https://github.com/Devetak/ESRIcascade.jl under the MIT License.
\end{abstract}

\section*{Background and Motivation}

Socio-economic systems are inherently networked. The failure of individual components can cascade through the network links, ultimately impacting the functionality of the system as a whole \cite{haldane2011systemic}. This structural interdependence gives rise to systemic risk.

A prominent example of the materialization of such risk is the 2008 financial crisis. There, the distress of a single institution within a financial network of interbank lending initiated cascading failures, which then resulted in the Great Financial Crisis. In response, researchers developed quantitative frameworks to model and estimate the vulnerability of networks to such cascading failures. A key conceptual paradigm shift from this line of work was the recognition that institutions may be ``too central to fail'' rather than merely ``too big to fail,'' a concept operationalized by network-based metrics such as DebtRank \cite{battiston2012debtrank}.

A similar logic applies to production networks and supply chains. Those networks comprise firms that produce by purchasing the outputs of other firms and whose output is itself used in other production processes. It is not surprising that, in these networks as well, a localized disruption can have large macroeconomic consequences. This is illustrated by the 2011 Great East Japan Earthquake and the subsequent tsunami and the Fukushima nuclear disaster. While the initial physical shock was geographically localized, its economic repercussions propagated along supply networks \cite{carvalho2021supply}. Notably, Inoue and Todo \cite{inoue2019firm} demonstrated that the indirect macroeconomic impact on firms not directly damaged by the disaster accounted for 2.3\% of GDP, a figure more than 100 times larger than the direct physical losses. This points to the existence of specific nodes in the supply chain networks whose disruption can impact an entire economy, regardless of the physical size of the initial shock.

To systematically quantify these vulnerabilities, the Economic Systemic Risk Index (ESRI) was proposed \citep{diem2022esri}. ESRI evaluates systemic importance by simulating the temporary failure of individual firms. A target firm is assumed to be temporarily inactive, demanding zero inputs from its suppliers and delivering zero outputs to its customers. The resulting supply and demand changes are then propagated through multiple tiers of the network. A firm's ESRI is then defined as the fraction of total national output that is expected to be lost or disrupted as a direct consequence of its failure.

When applied to empirical, nationwide firm-level supply networks, the ESRI reveals a small systemic risk core. For Hungary, this core consisted of only 32 companies, with each firm exhibiting an ESRI of approximately 23\% of national economic output \citep{diem2022esri}. Consequently, in the hypothetical scenario of an unmitigated failure, any single firm in this core would immediately jeopardize nearly a quarter of the national economy. Firm size was found to be a poor predictor of systemic importance; while several critical nodes were large enterprises, a significant portion of them were relatively small firms situated in highly central structural positions within the network topology. Similar structural systemic cores have since been documented across specific supply networks in various other national contexts \citep{zelbi2026mitigation}.

The ESRI framework has since been used in a range of empirical and methodological applications, including the monitoring of supply networks from mobile-phone data \citep{reisch2022mobile}, the study of information loss caused by network aggregation \citep{diem2024predictability}, financial-stability analysis \citep{tabachova2024financial, fialkowski2026data}, rapid decarbonization scenarios \citep{stangl2024decarbonization}, and the analysis of changes in firm-level systemic risk over time \citep{mancini2026evolution}. It has also become an important validation target for methods that reconstruct firm-level production networks from incomplete or aggregated data \citep{bacilieri2023partial,fessina2026inferring,devetak2026industry}, and has been used to study how supply-chain rewiring can alter or mitigate systemic risk \citep{reisch2026rewiring,zelbi2026mitigation}. These developments form part of a broader effort to map and analyze firm-level supply networks \citep{pichler2023alliance}, and ESRI-based results have also informed governmental work on supply-chain risk and resilience \citep{governmentofficeforscience2026globalsupplychains}.

Computationally and conceptually, what ESRI does is ask: if this firm were to close, how much of the economy would be impacted in the short term? To compute the ESRI of an entire economy means asking this question over and over again for each firm in it. Since, for each firm, one needs to propagate the shock of closing down across a large network, these individual calls are not cheap. \texttt{ESRIcascade.jl} addresses these limitations by making individual calls cheaper without changing the underlying algorithm.

Existing ESRI implementations reproduce the model but become costly for repeated firm-level analyses of large networks. \texttt{ESRIcascade.jl} provides a research implementation focused on reducing runtime and memory use without changing the ESRI model.

\section*{Software Overview}

Let \(W_{ij}\) denote the value supplied by firm \(i\) to firm \(j\). Firms must have consecutive one-based indices, and the industry identifier for each firm must follow the same ordering. A simple analysis would be:

\begin{lstlisting}[style=julia]
using ESRIcascade
using SparseArrays

supplier = Int.(transactions.supplier)
customer = Int.(transactions.customer)
value = Float64.(transactions.value)
firm_industry = Int.(firms.industry_id)

N = length(firm_industry)
W = sparse(supplier, customer, value, N, N)

# Julia uses column-major (column-first) indexing.
# W[i, j] is the transaction from supplier i to customer j.
# input_classification is a K x K matrix in the same industry order:
# rows are supplying industries and columns are customer industries.
# 0 = no short-run downstream effect;
# 1 = non-essential input; 2 = essential input.
info = IndustryInfo(firm_industry, input_classification)
econ = ESRIEconomy(W, info)
\end{lstlisting}

When firms are assigned IHS industry codes, the package also provides the corresponding industry-code order and a bundled \(616\times616\) input classification derived from the processed IHS industry survey data in the replication archive of Pichler et al. \citep{pichler2022pandemic,pichler2022zenodo}:

\begin{lstlisting}[style=julia]
codes = ihs_industry_codes()
industry_id = Dict(code => i for (i, code) in pairs(codes))
firm_industry = [industry_id[c] for c in firms.industry_code]
info = IndustryInfo(firm_industry, ihs_input_classification())
\end{lstlisting}

A standard run closes each firm in turn and returns one score per firm. A single firm can also be selected using its index:

\begin{lstlisting}[style=julia]
scores = esri(econ)
single_score = esri(econ, 17) # only compute the ESRI of firm 17
\end{lstlisting}

More granular scenarios can be represented with a capacity vector \(\psi\).
Here, one denotes normal operation, zero denotes closure, and intermediate
values denote partial capacity:

\begin{lstlisting}[style=julia]
psi = ones(N)
psi[17] = 0.0
psi[23] = 0.5
scenario = esri_shock(econ, psi)
\end{lstlisting}

The ESRI score of a firm is the percentage of the economy that is shut down by the firm shutting down. Sometimes it is of interest to compute different percentages, like emissions or workers. For that purpose, alternative weights can be used instead:

\begin{lstlisting}[style=julia]
worker_weights = Float64.(firms.workers)
worker_weighted_scores = esri(econ; final_weights = worker_weights)
\end{lstlisting}

Usually the upstream and downstream propagation channels are combined. In \texttt{ESRIcascade.jl}, we enable the user to select a single propagation channel; this distinguishes losses transmitted through suppliers from losses transmitted through customers:

\begin{lstlisting}[style=julia]
scores_min = esri(econ; combine = :min)  # default
scores_upstream = esri(econ; combine = :upstream)
scores_downstream = esri(econ; combine = :downstream)
\end{lstlisting}

\section*{Algorithmic Optimizations}

In this section, we present the main differences between \texttt{ESRIcascade.jl} and the legacy C++/R code from \cite{fastcascade, esri_tutorial}. We aim to explain the algorithmic improvements we propose and why they matter.

The problem of computing the ESRI of an entire economy is embarrassingly parallel. Hence, we focused our attention on making the ESRI computation for a single firm-closure scenario faster.

Essentially, all the major optimizations we perform boil down to a simple idea. Instead of representing ESRI as a series of matrix--matrix and matrix--vector multiplications, we unroll these operations by traversing the production network directly. This traversal exposes ESRI-specific structures that we can exploit to avoid redundant computations.

The computation of ESRI is split into two parts. The upstream propagation step computes how a reduction in a firm's output reduces the demand faced by its suppliers. This step is linear and relatively fast to compute. The downstream propagation step instead accounts for production that cannot take place because firms lose deliveries of necessary inputs. This step is more complex because the losses are not proportional, but are instead mediated by a generalized Leontief production function. It therefore takes longer to compute, but also provides more opportunities for optimization.

One optimization we perform is to store each ESRI-step operator in a sparse format aligned with the direction in which it is traversed. This makes memory access more contiguous and avoids processing inactive links and zero entries. It can also make the tight accumulation loops easier to vectorize, although the main gain comes from fewer operations and less memory traffic rather than from SIMD alone.

For every customer firm, the generalized Leontief production function needs to know how much input remains available from each supplier industry. \texttt{fastcascade} calculates these quantities by constructing a dense \(N \times M\) matrix, where \(N\) is the number of firms and \(M\) is the number of industries. It first scales and multiplies sparse matrices, then extracts the relevant industry columns and applies the essential minima and non-essential sums.

We replace this with an implementation that evaluates the same equation from the opposite direction. For each supplier, we traverse the observed supplier--customer links and add lost supply directly to the appropriate customer accumulator. After all affected links have been visited, the customer's essential constraint is obtained from its largest industry-specific loss, while its non-essential constraint is obtained from its total non-essential loss. We also directly accumulate the linear parts of the production function. This is faster because each active link is visited once and its contribution is written directly to the final accumulator. It avoids constructing and repeatedly scanning dense \(N \times M\) intermediates, thereby reducing both arithmetic and memory traffic.

For a given customer industry, it is convenient and economically meaningful to assign each input industry a specific type. Essential inputs are inputs without which a firm is unable to produce, such as nails for a carpenter. These enter the Leontief part of the production function. Non-essential inputs can be important but are not central to production. An example is food supplied to a factory's cafeteria. These enter the production function through its linear part. A third category, no short-run impact, contains inputs that have no short-term effect on a firm. Consulting services are one example. These links are omitted from the downstream propagation.

\section*{Validation and Performance}

Since we built \texttt{ESRIcascade.jl} to be equivalent to the C++/R implementations \cite{fastcascade, esri_tutorial}, we ensure this by running ESRI computations on the same economy using both packages. We report that the values match up to the convergence tolerance in all our tests.

Since our aim is to speed up ESRI computations, we benchmark \texttt{ESRIcascade.jl}. We do so by using a synthetic power-law network, which mirrors real-life networks. We assign firms evenly across the 616 IHS industries and compute the ESRI of every firm. We test different production functions: a linear production function, a mixed production function like the one used in the original \cite{diem2022esri}, and an IHS production function that also makes use of non-essential input classifications from \citep{pichler2020restart,pichler2022pandemic}. We tested these configurations with 1 and 4 threads. We ran all the experiments on economies of 10,000 firms. Additionally, for the realistic IHS production function, we ran an experiment with 100,000 firms. We note that 100,000 firms is a realistic use case for the ESRI algorithm.

\begin{table}[htbp]
\centering
\caption{Runtime comparison between \texttt{ESRIcascade.jl} and \texttt{fastcascade}.}
\label{tab:runtime-comparison}
\renewcommand{\arraystretch}{1.15}
\setlength{\tabcolsep}{9pt}

\begin{tabular}{llcrrr}
\toprule
\textbf{Size} &
\textbf{Production Function} &
\textbf{Threads} &
\textbf{\texttt{ESRIcascade.jl}} &
\textbf{\texttt{fastcascade}} &
\textbf{Speedup} \\
&
&
&
\multicolumn{2}{c}{\textit{ms per firm}} &
\\
\cmidrule(lr){4-5}

\multirow{6}{*}{\textbf{10k}}
& \multirow{2}{*}{IHS}
    & 1 & 0.777 & 172.422 & 222$\times$ \\
&   & 4 & 0.296 &  61.432 & 207$\times$ \\
\cmidrule(lr){2-6}

& \multirow{2}{*}{Legacy}
    & 1 & 1.103 & 210.009 & 190$\times$ \\
&   & 4 & 0.343 &  72.083 & 210$\times$ \\
\cmidrule(lr){2-6}

& \multirow{2}{*}{Linear}
    & 1 & 0.826 & 250.429 & 303$\times$ \\
&   & 4 & 0.293 &  86.772 & 295$\times$ \\

\midrule

\multirow{2}{*}{\textbf{100k}}
& \multirow{2}{*}{IHS}
    & 1 & 9.885 & 1900.390 & 192$\times$ \\
&   & 4 & 4.222 &  587.493 & 139$\times$ \\

\bottomrule
\end{tabular}
\end{table}

Table \ref{tab:runtime-comparison} presents the results. We note that for smaller networks the speedup reaches a factor of $300$ in the linear case, which is expected given the care we took in optimizing the linear accumulation. For the 100,000-firm network, the factors are slightly lower but nevertheless lead to significant time savings. For example, the \texttt{ESRIcascade.jl} code for the example using 4 threads ran for 7 minutes, while the legacy code took slightly more than 16 hours. We think this speedup will be welcomed by researchers working with ESRI. In these experiments, we also tracked maximum RAM usage. Since a lot of the economic data is duplicated across workers in the legacy code, we find that the extra data usage matches expectations. In our case, since we keep most of the economic data as a single struct, this is not the case. This might be relevant for larger economies when running multiple parallel computations. For reference, a single-threaded computation of 100,000 firms uses 1.1 GB of RAM, while a four-threaded one consumes 2.5 GB. For the \texttt{fastcascade} code, the numbers are 2.1 GB and 7.6 GB, respectively.

\section*{Conclusion}

This work presents \texttt{ESRIcascade.jl}, a Julia package for computing the Economic Systemic Risk Index. The package preserves the ESRI model while replacing costly matrix-based operations with direct traversal of sparse production-network links. In our benchmarks, \texttt{ESRIcascade.jl} is up to 300 times faster than the legacy implementation. These improvements make repeated analyses of larger production networks more practical and may support broader use of ESRI in empirical and methodological research.

\newpage
\bibliographystyle{unsrtnat}
\bibliography{paper}
\end{document}